\documentclass[aps,pre,twocolumn,reprint]{revtex4-2}
\usepackage{amsmath,amssymb}
\usepackage{graphicx}

\begin{document}

\title{Nonsingular increase in magnetic susceptibility and transition in
  universality in site-diluted Ising model in two dimensions}
\author{Eduardo C. Cuansing}
\email{eccuansing@up.edu.ph}
\affiliation{Institute of Mathematical Sciences and Physics, University of
  the Philippines Los Ba\~{n}os, Laguna 4031, Philippines}

\date{\today}

\begin{abstract}
  We study the effects of dilution to the critical properties of
  site-diluted Ising model in two dimensions using Monte Carlo simulations.
  Quenched disorder from the dilution is incorporated into the Ising model
  via random empty sites on the square lattice of Ising spins.
  Thermodynamic quantities such as the magnetization $M$ per spin, energy
  $E$ per spin, magnetic susceptibility $\chi$ per spin, and specific heat
  $C$ per spin are then calculated after the system has equilibrated.
  At small dilution concentrations $d<0.1$, we find that the value of the
  critical exponent $\beta$ does not deviate from its pure Ising value.
  At higher dilution concentrations $d>0.1$, however, we find $\beta$ to
  strongly depend on the value of $d$. We are able to locate a critical
  temperature $T_c$ and a critical dilution concentration $d_c$ where the
  phase transition occurs. We find $T_c$ to depend linearly on $d$. In the
  phase diagrams of $M$, $E$, $\chi$, and $C$, we find that the phase
  transition line eventually disappears at high dilutions. Our results
  suggest that there is a transition from Strong Universality at low
  dilution to Weak Universality at high dilution. Lastly, we find a wide
  and nonsingular increase in the magnetic susceptiblity $\chi$ at the low
  temperature and high dilution region. 
\end{abstract}

\maketitle

\section{Introduction}
\label{sec:intro}

The interplay between disorder, particle interactions, and thermal
fluctuations can lead to interesting physics in processes such as the
non-Fourier transport of heat in nanosystems \cite{benenti2023}, the
localization of particles in disorder-driven many-body systems
\cite{geissler2020}, the melting of magnetic vortices in high-temperature
superconductors \cite{goldschmidt2005}, and in phase transitions in spin
systems \cite{lima2015,delfino2017,vatansever2020}. The addition of quenched
disorder in the occupation of lattice sites in classical Ising spin systems,
in particular, may lead to different critical properties from the pure
system. The Harris criterion \cite{harris1974} states that the value of the
specific heat critical exponent $\alpha_{pure}$ of the pure Ising system
indicates whether the introduction of disorder is relevant, when
$\alpha_{pure}>0$, or irrelevant, when $\alpha_{pure}<0$. This criterion has
been confirmed in site-diluted Ising systems in three dimensions
\cite{ballesteros1998,calabrese2003,hasenbusch2007}. In two dimensions,
however, $\alpha_{pure}=0$ and the problem is outside the scope of the Harris
criterion. Two possible scenarios have been proposed when dilution is
introduced in Ising systems in two dimensions. In the Weak Universality
hypothesis, the addition of dilution does not change the form of the
power-law scaling relations \cite{kim1994a}. However, the values of the
critical exponents vary depending on the amount of dilution, although
critical exponent ratios such as $\beta/\nu$ and $\gamma/\nu$ remain the same
\cite{kim1994a,fahnle1992,kim1994b,kuhn1994,hadjiagapiou2008,schrauth2018}. 
On the other hand, the Strong Universality hypothesis states that the
introduction of dilution does not change the values of the critical
exponents. Instead, dilution leads to, at most, logarithmic corrections to
the critical scaling relations of the system
\cite{dotsenko1983,shalaev1994,ballesteros1997,selke1998,kenna2008,zhu2015}. 

In this work, we revisit site-diluted Ising model in two dimensions. We
investigate the effects of introducing dilution at concentrations that range
from perturbatively small to high nonperturbative values. In
Section~\ref{sec:model}, we introduce site-diluted Ising model in two
dimensions and show the details on how quantities such as the magnetization
$M$ per spin or the magnetic susceptibility $\chi$ per spin are calculated
from our Monte Carlo simulations. The results of our simulations are
discussed in Section~\ref{sec:results}. Here, we show how thermodynamic
quantities behave as the temperature and the dilution concentration are
varied. We locate a critical temperature $T_c$ and a critical dilution
$d_c$ where the phase transition occurs. We also find how $T_c$ changes as
the dilution is varied. Furthermore, data on how the critical exponent
$\beta$ changes with $d$ suggests a transition from Strong Universality at
low dilution to Weak Universality at high dilution. Lastly, we show that
the magnetic susceptibility increases to a nonsingular maximum at low
temperatures and high dilution. The summary and conclusion is discussed in
Section~\ref{sec:conclusion}. 

\section{Site-diluted Ising model}
\label{sec:model}

In site-diluted Ising model on the square lattice, sites on the lattice
are either occupied or empty according to the site occupation probability
$p = 1-d$, where $d$ is the dilution concentration. The random occupation
of sites are uncorrelated and each occupied site $i$ contains an Ising spin
$\sigma_i$ which has two possible orientations, either up ($\sigma_i = +1$)
or down ($\sigma_i = -1$). The Hamiltonian for the model is
\begin{equation}
  H = -J\sum\limits_{\left\langle i,j\right\rangle}\epsilon_i \epsilon_j
  \sigma_i \sigma_j + h\sum\limits_i \epsilon_i \sigma_i,
  \label{eq:hamiltonian}
\end{equation}
where the $\epsilon_i$ are the site-occupation variables and are either
$1$ (occupied site) or $0$ (empty site), $J>0$ is the ferromagnetic
coupling constant, $h$ is the applied external magnetic field, and the
sum under $\left\langle i,j\right\rangle$ are over all nearest-neighbor
sites $i$ and $j$. In this work, we study the properties of the system
without the applied external magnetic field and thus, set $h=0$. A
realization of the disordered configuration of occupied sites is quenched
at the outset of our Monte Carlo simulation.

We implement the Metropolis algorithm with single-spin flips
\cite{newman1999} in our Monte Carlo simulations to determine the
equilibrium properties of the disordered spin systems. Once the process has
equilibrated, we determine the total energy and magnetization of the system
after every Monte Carlo step as
\begin{align}
  & E_{MC} = -J\sum\limits_{\left\langle i,j\right\rangle}\epsilon_i \epsilon_j
  \sigma_i \sigma_j,\label{eq:emc}\\
  & m = \sum\limits_i \sigma_i.\label{eq:m}
\end{align}
The simulation is then repeated several times, each time with a different
realization of the disorder due to the dilution. At the end of the
simulation, the energy $E$ per site, magnetization $M$ per site, magnetic
susceptibility $\chi$ per site, and specific heat $C$ per site are
calculated as
\begin{align}
  & E = \left[\frac{\left\langle E_{MC}\right\rangle}{2 L^2}\right]_r,
  \label{eq:energy}\\
  & M = \left[\frac{\left\langle m\right\rangle}{L^2}\right]_r,
  \label{eq:magnetization}\\
  & \chi = \left[\frac{\left\langle m^2\right\rangle - \left\langle
      m\right\rangle^2}{k_B T L^2}\right]_r,
  \label{eq:chi}\\
  & C = \left[\frac{\left\langle E_{mc}^2\right\rangle-\left\langle E_{mc}
      \right\rangle^2}{k_B T^2 L^2}\right]_r,
  \label{eq:spheat}
\end{align}
where the angled brackets mean the average over all Monte Carlo steps, the
$\left[\ \right]_r$ means the average over disorder realizations,
$k_B$ is the Boltzmann constant, $T$ is the temperature, and $L\times L$ is
the size of the lattice.

The pure, undiluted, Ising model on the square lattice is known to undergo
a continuous phase transition \cite{kramers1941} at the critical temperature
$T_c = 2.27$, in units of $J/k_B$. At temperatures around $T_c$, the pure
Ising model follows scaling relations
\begin{align}
  & M \sim \left|t\right|^{\beta},\label{eq:mscaling}\\
  & \chi \sim \left|t\right|^{-\gamma},\label{eq:chiscaling}\\
  & C \sim \left|t\right|^{-\alpha},\label{eq:spheatscaling}
\end{align}
where $t \equiv \left(T - T_c\right)/T_c$ and the critical exponents have
values $\beta = 1/8 = 0.125$, $\gamma = 7/4 = 1.75$, and $\alpha = 0$ in
the square lattice \cite{newman1999}. When dilution is
introduced into the model, the Strong Universality hypothesis states that
the dilution does not change the values of the critical exponents and,
at most, only makes logarithmic corrections to the scaling
\cite{dotsenko1983,shalaev1994,ballesteros1997,selke1998,kenna2008,zhu2015}.
On the other hand, in the Weak Universality hypothesis scenario, the
power-law scaling relations remain the same but with the values of the
critical exponents strongly depending on the amount of dilution
\cite{fahnle1992,kim1994a,kim1994b,kuhn1994,hadjiagapiou2008,schrauth2018}. 

\section{Numerical results}
\label{sec:results}

We investigate the equilibrium properties of site-diluted Ising spin systems
under various dilution concentration and temperature values. The square
lattices have sizes $200\times 200$ and sites at the boundaries are fixed as
empty. There are $20000$ Monte Carlo steps, with the first $5000$ transient
steps discarded. The process is then repeated for a different realization
of the disorder. In total, there are $100$ realizations of the disorder for
each dilution concentration value.

\begin{figure}[h!]
  \centering
  \includegraphics[width=0.98\columnwidth,clip]{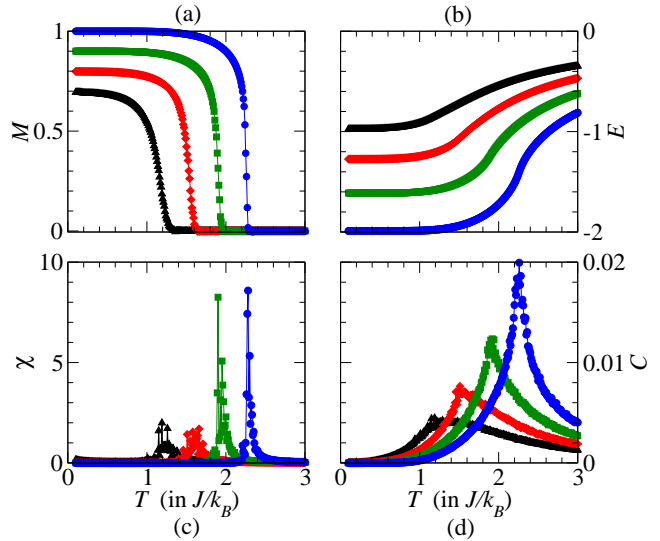}
  \caption{Plots of the (a) magnetization $M$ per site, (b) the total
    energy $E$ per site, (c) the magnetic susceptibility $\chi$ per site,
    and (d) the specific heat $C$ per site as functions of the temperature
    $T$. Dilution concentrations are $d = 0$ (blue circles), $d = 0.1$
    (green squares), $d = 0.2$ (red diamonds), and $d = 0.3$ (black
    triangles). Lattices are of size $200 \times 200$ with $100$
    realizations of the dilution disorder. The lines are drawn to aid
    the eye.}
  \label{fig:fn_T}
\end{figure}

Shown in Fig.~\ref{fig:fn_T} are plots of the magnetization $M$ per site,
total energy $E$ per site, magnetic susceptibility $\chi$ per site, and
specific heat $C$ per site as the temperature is varied from $T=0.1$ to
$T=3$ (in this work, all temperature values are in units of $J/k_B$) in
steps of $\Delta T=0.01$ for four values of the dilution concentration,
$d=0,0.1,0.2$, and $0.3$. In the figure, data for the undiluted, $d=0$,
system indicates a continuous phase transition at around $T\approx2.26$.
As the amount of dilution is increased, the transition temperature decreases
to lower temperatures. For large dilution concentrations, the curves for
$M$ and $C$ approach the transition temperature with different slopes. In
addition, in Fig.~\ref{fig:fn_T}(c), at high dilution, hints of a
nonsingular maximum in the specific heat at temperatures $T>T_c$ is
visible, which was also found previously by Selke, et al\cite{selke1998}.

\begin{figure}[h!]
  \centering
  \includegraphics[width=0.98\columnwidth,clip]{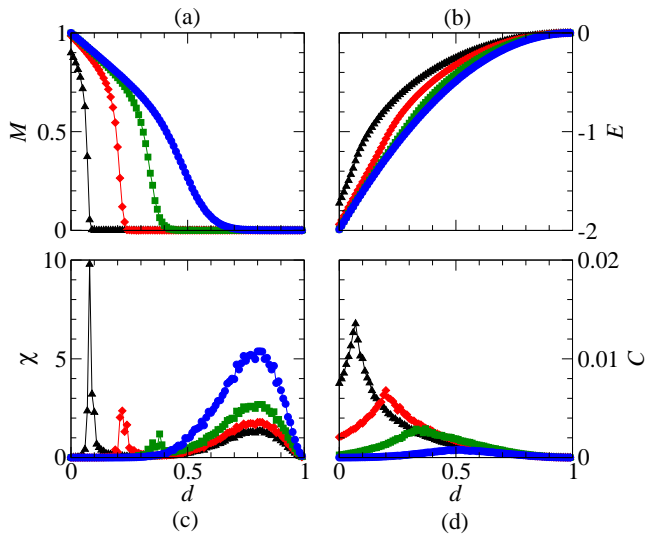}
  \caption{Plots of (a) the magnetization $M$ per site, (b) the total
    energy $E$ per site, (c) the magnetic susceptibility $\chi$ per site,
    and (d) the specific heat $C$ per site as functions of the dilution
    concentration $d$. The temperatures are $T = 0.5$ (blue circles),
    $T = 1$ (green squares), $T = 1.5$ (red diamonds), and $T = 2$ (black
    triangles), in units of $J/\rm{k}_B$. Data are from $200\times 200$
    lattices with $100$ realizations of the disorder due to the dilution.}
  \label{fig:fn_d}
\end{figure}

Shown in Fig.~\ref{fig:fn_d} are plots of $M$, $E$, $\chi$, and $C$ as the
dilution concentration $d$ is varied from $d=0$ to $d=1$, in steps of
$\Delta d=0.01$, for four values of the temperature, $T=0.5,1,1.5$, and
$2$, which are less than the critical temperature $T_c=2.27$ of the pure
Ising system. The plots suggest the occurence of a continuous phase
transition, at a critical dilution $d_c$, as $d$ is increased from $d=0$
to $d=1$. We can see that the value of $d_c$ decreases as the temperature
is increased, implying that the transition from the ferromagnetic state to
the paramagnetic state requires less dilution when the temperature is high.
In addition, at low temperatures, Fig.~\ref{fig:fn_d}(c) shows wide but
nonsingular increase in the susceptibility $\chi$ at high dilution. This
increase in $\chi$ can be attributed to high fluctuations between isolated
islands of spins, instead of fluctuations between individual spins, in
systems with high dilution and at low temperatures.

\begin{figure}[h!]
  \centering
  \includegraphics[width=0.97\columnwidth,clip]{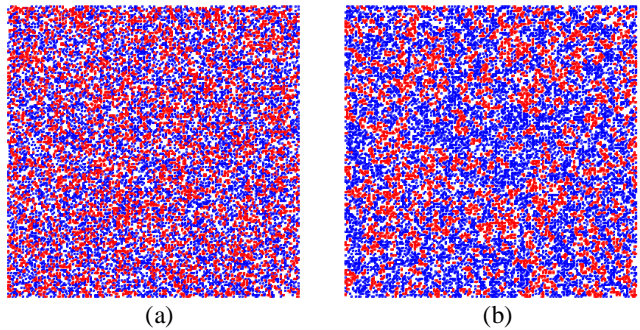}
  \caption{Sample configurations of the spins in $200\times 200$ diluted
    lattices. Up spins are blue, down spins are red, and empty sites are
    white. System temperatures are (a) $T = 2$ and (b) $T = 0.5$, in units
    of $J/\rm{k}_B$. Both systems have the same dilution concentration
    $d = 0.6$.}
  \label{fig:sample}
\end{figure}

For comparison, we show in Fig.~\ref{fig:sample} sample configurations of
the diluted systems when $d=0.6$ for a system at a high temperature $T=2$
and at a low temperature $T=0.5$. There are two competing effects in the
Ising model. The nearest-neighbor interaction between spins encourages
neighboring spins to align while temperature provides spins with the means
to flip, thereby leading to thermal fluctuations. At high temperatures,
thermal fluctuations are large and many nearest-neighboring spins would
point in opposite directions. Dilution does not affect thermal fluctuations
but the presence of empty sites in the system can lead to higher $\chi$.
This is because high dilution will separate the system into isolated
islands of net spins. At high temperatures, as shown in
Fig.~\ref{fig:sample}(a), these islands will most likely have only a small
or zero net spin because of thermal fluctuations. At low temperatures, as
shown in Fig.~\ref{fig:sample}(b), neighboring spins within an island will
likely align, which would lead to a system of isolated islands each with net
nonzero spins. However, because the interaction is only between
nearest-neighboring spins and there are empty sites between islands, the
islands do not interact with each other. Since $\chi$ is a measure of
spin fluctuations, see Eq.~(\ref{eq:chi}), net spin fluctuations among
isolated islands will lead to large values in $\chi$.

\begin{figure}[h!]
  \centering
  \includegraphics[width=0.48\columnwidth,clip]{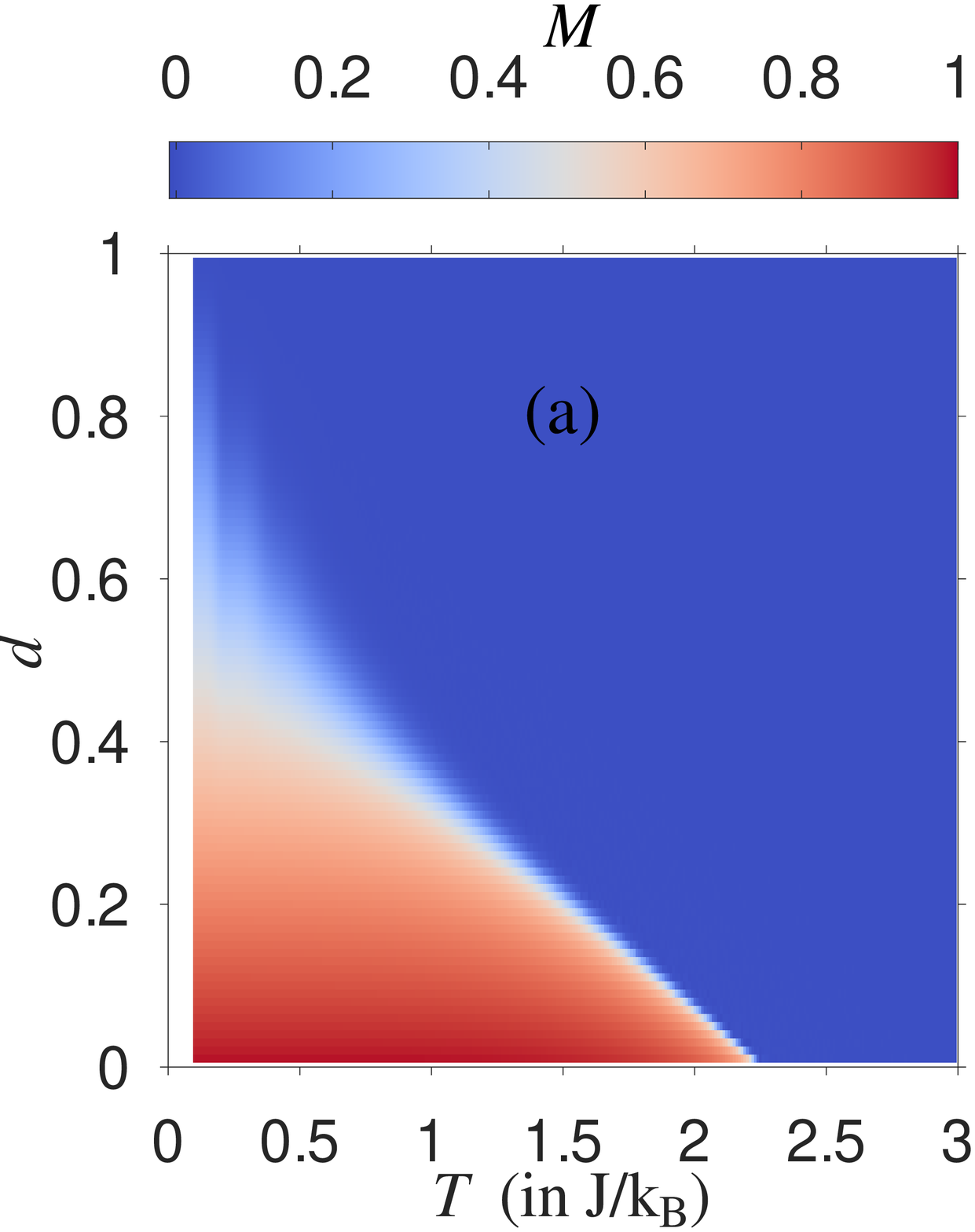}~
  \includegraphics[width=0.48\columnwidth,clip]{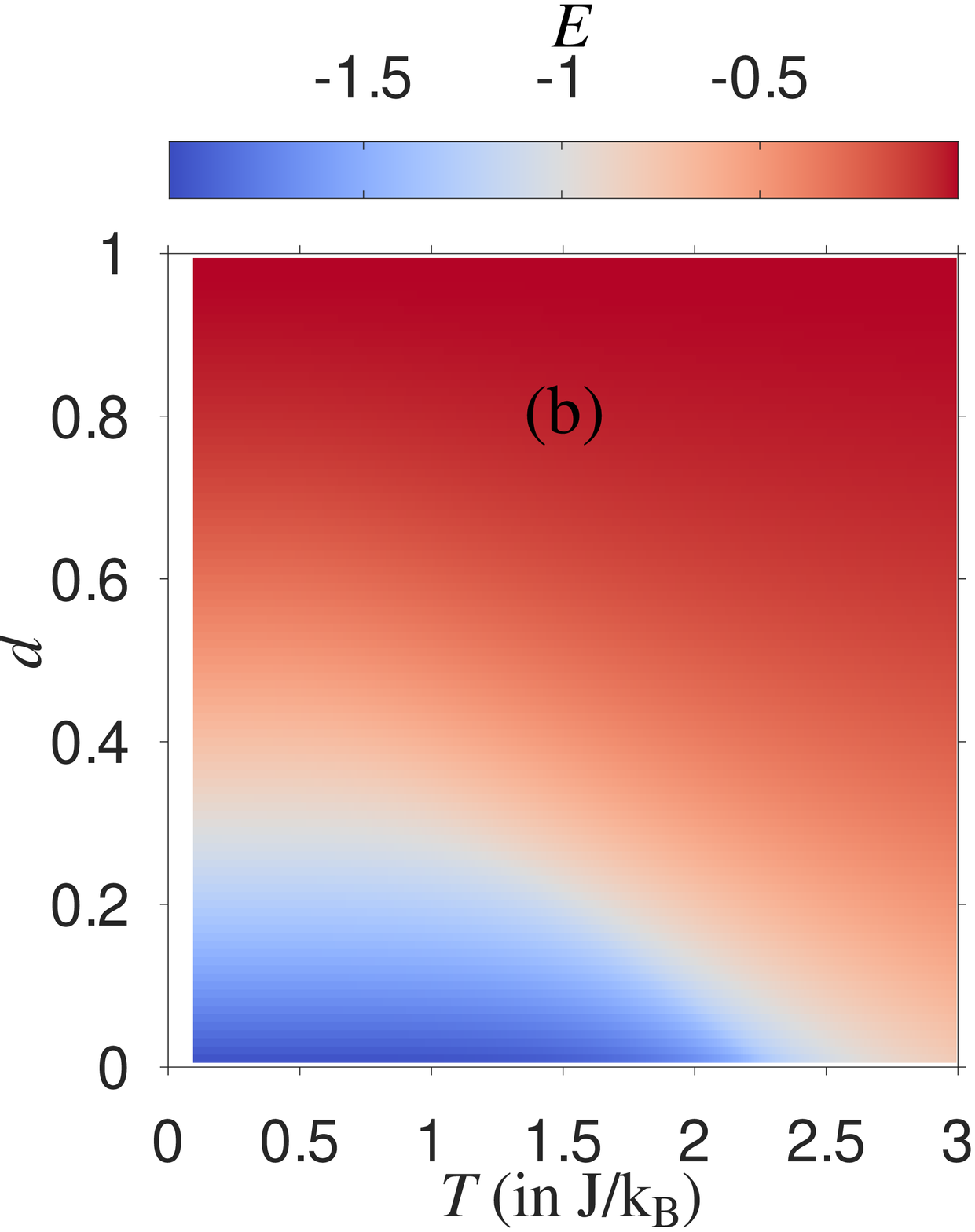}\\
  \vspace*{8pt}
  \includegraphics[width=0.48\columnwidth,clip]{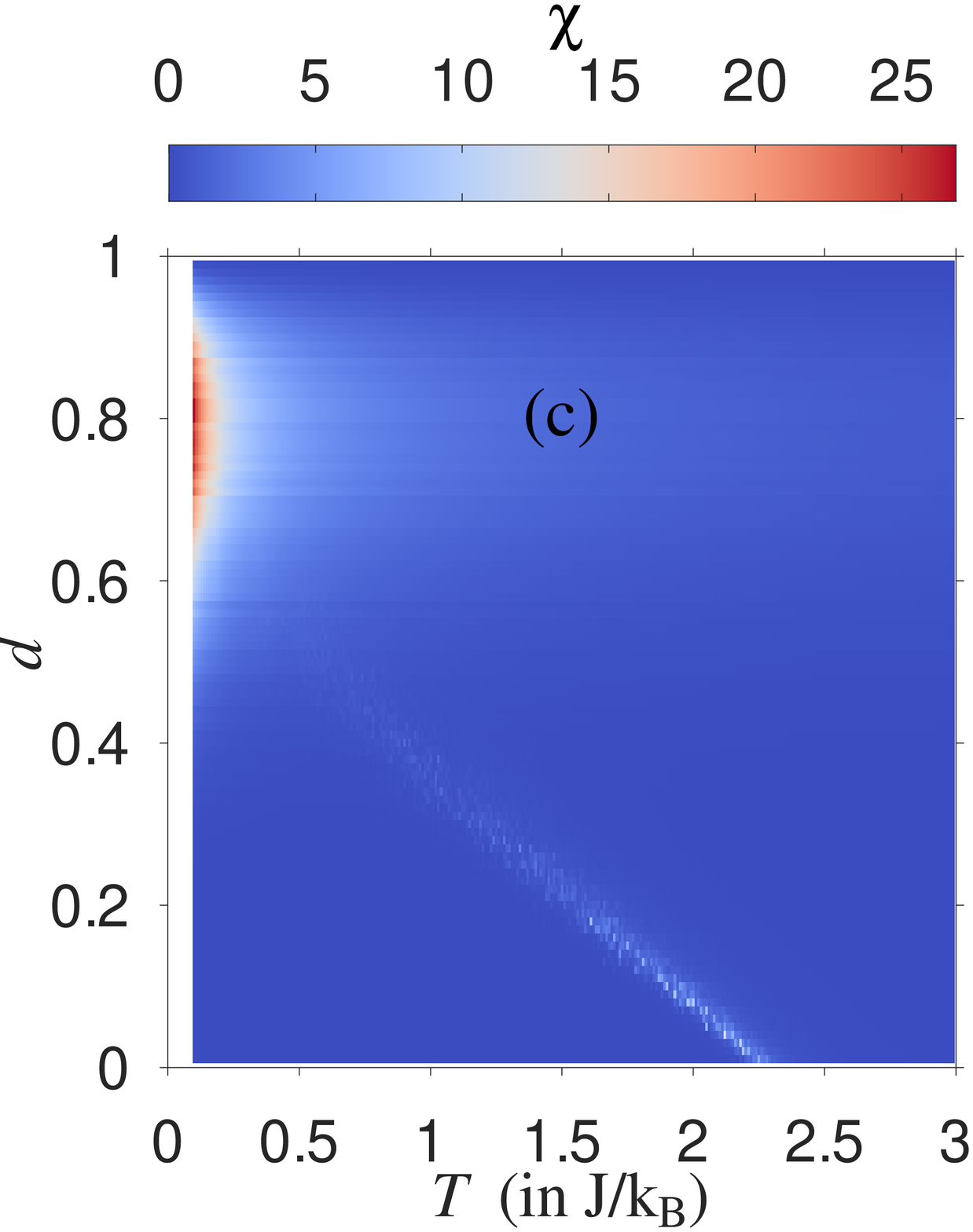}~
  \includegraphics[width=0.48\columnwidth,clip]{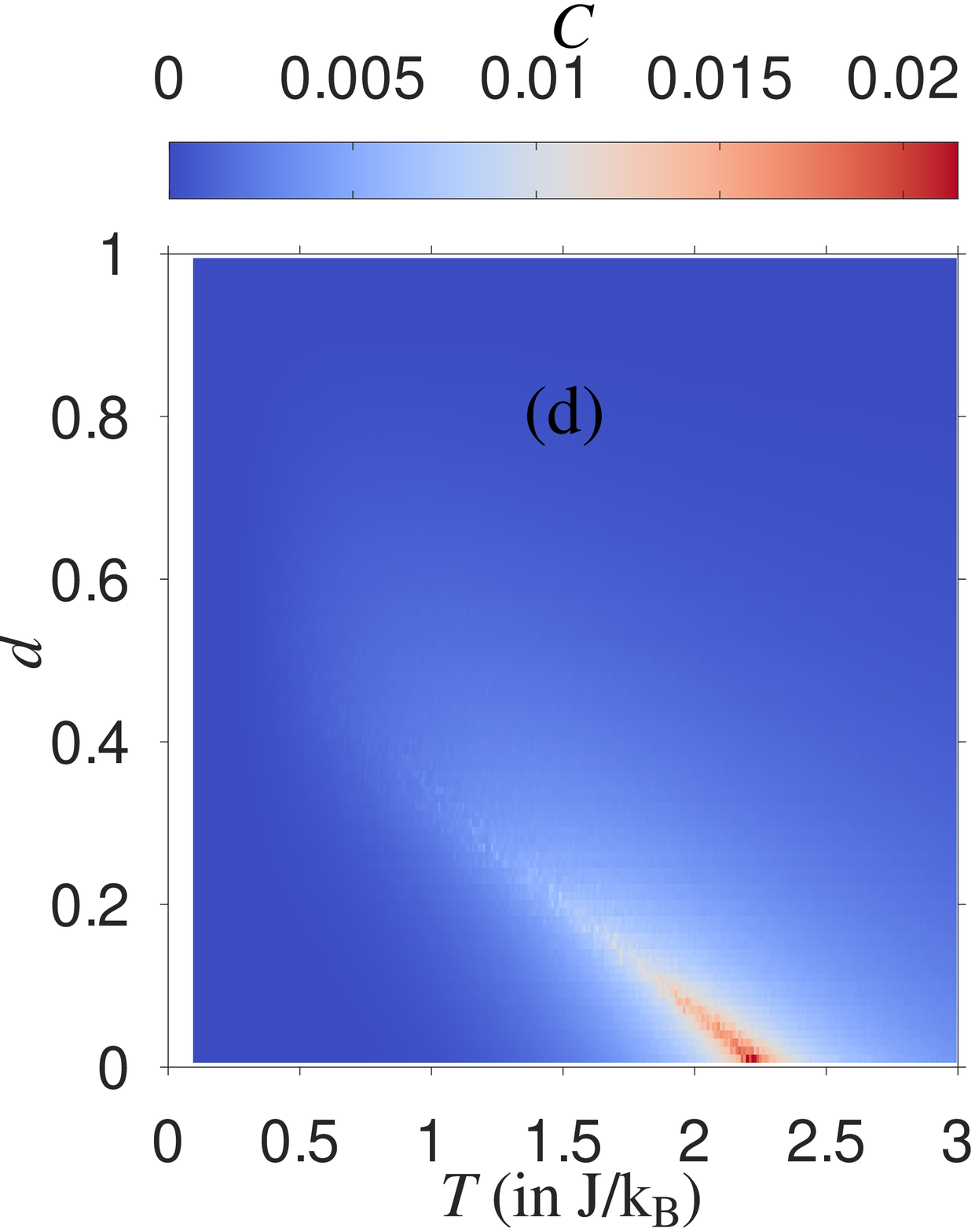}\\
  \caption{Phase diagrams of (a) the magnetization $M$ per site,
    (b) the total energy $E$ per site, (c) the magnetic susceptibility
    $\chi$ per site, and (d) the specific heat $C$ per site. Data are
    from lattices of size $200\times 200$ with $100$ realizations of the
    dilution disorder.}
  \label{fig:phasediagrams}
\end{figure}

The full picture can be seen in the phase diagrams shown in
Fig.~\ref{fig:phasediagrams}. In Fig.~\ref{fig:phasediagrams}(a) where
$M$ is shown, there is a sharp transition line at high $T$ and low $d$ that
eventually dissolves away at low $T$ and high $d$. This transition line
coincides with the region where $E$ changes value, as shown in
Fig.~\ref{fig:phasediagrams}(b). The transition line also shows up in the
diagrams for $\chi$, in Fig.~\ref{fig:phasediagrams}(c), and $C$, in
Fig.~\ref{fig:phasediagrams}(d). The eventual disappearance of the
sharp transition line at low $T$ and high $d$ in specific heat, in particular,
is consistent with what was previously reported in \cite{kim1994b}. Also, the
wide and nonsingular increase in $\chi$ at the low $T$ and high $d$ region
appears in Fig.~\ref{fig:phasediagrams}(c).

\begin{figure}[h!]
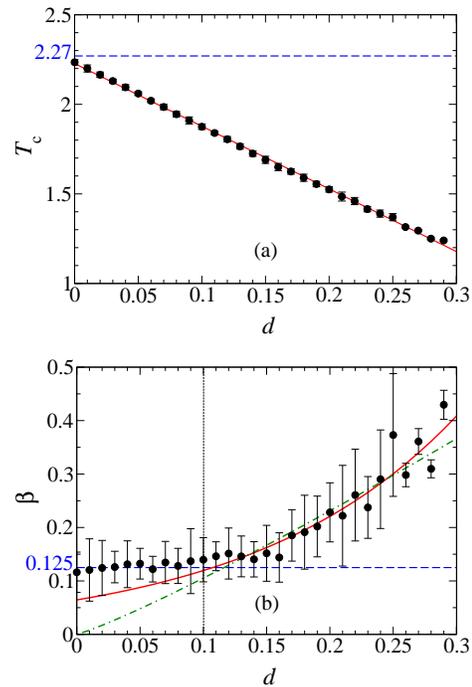

  \centering
  \includegraphics[width=0.7\columnwidth,clip]{fig_5a.eps}\\
  \vspace*{8pt}
  \includegraphics[width=0.7\columnwidth,clip]{fig_5b.eps}
  \caption{Plots of (a) the critical temperature $T_c$ as a function of the
    dilution concentration $d$ and (b) the critical exponent $\beta$ as a
    function of $d$. Also shown in (b) are the exponential fits (red line)
    and the power-law fits (green dash-dot line). The vertical line in
    (b) separates the plot into $d<0.1$ and $d>0.1$ regions.}
  \label{fig:beta}
\end{figure}

The data for $M$ can be used to determine how the critical exponent $\beta$,
appearing in Eq.~(\ref{eq:mscaling}), behaves as the dilution concentration
$d$ and the temperature $T$ are varied. To determine $\beta$ for a given
value of $d$, we first locate an estimate critical temperature $T_c$ value
that best fits, with a correlation coefficient $|R|>0.99$, the power-law
scaling relation in Eq.~(\ref{eq:mscaling}). Shown in Fig.~\ref{fig:beta}(a)
is how the estimate $T_c$ behaves as $d$ is increased from $d=0$ to $d=0.3$.
Notice that for the undiluted $d=0$ system we expect a critical temperature
$T_c = 2.27$. The slight difference with our estimated $T_c$ can be
attributed to finite-size effects \cite{fisher1972}. A linear fit
\begin{align}
  T_c = -3.49811~d + 2.22756
  \label{eq:Tc_d}
\end{align}
with a correlation coefficient $|R|=0.99906$ indicates a linear dependence
of the estimated $T_c$ to the dilution concentration in the region from
$d=0$ to $d=0.3$, which is consistent, including the numerical values of the
slope and y-intercept, with what was previously found by Harris
\cite{harris1974}, up to the linear order in $d$. Notice that our data is
only up to $d=0.3$. A further quadratic dependence in $d$ might appear at
higher dilution $d>0.3$.

Shown in Fig.~\ref{fig:beta}(b) are the values of $\beta$ when $d$ is varied.
The dash line indicates the pure Ising value $\beta = 0.125$. For small
dilutions $d<0.1$, we find $\beta$ to be consistent with that for the
pure Ising system, thereby suggesting Strong Universality. However, as $d$
is further increased, the value of $\beta$ moves away from the pure Ising
system value and changes strongly with $d$, suggesting Weak Universality.
Making exponential and power-law fits to the values of $\beta$ when $d>0.1$,
we get
\begin{align}
  \beta & = \left(0.0645\right) e^{\left(6.1512\right) d},~~ |R| = 0.91895,
  \label{eq:beta_d_exp}\\
  \beta & = \left(1.4356\right) d^{\left(1.1347\right)},~~ |R| = 0.87621,
  \label{eq:beta_d_pow}
\end{align}
with their corresponding correlation coefficients $|R|$. The behavior of
$\beta$ as the dilution concentration is varied suggests a transition in
the critical behavior from Strong Universality at low dilution to Weak
Universality at high dilution.

The values of the critical exponents $\gamma$ and $\alpha$ in the diluted
systems can be determined from the scaling relations shown in
Eqs.~(\ref{eq:chiscaling}) and (\ref{eq:spheatscaling}), respectively. The
ratio of critical exponents, such as $\gamma/\nu$ and $\alpha/\nu$, can also
be determined from finite-size scaling \cite{binder1981}. However, our data
on $\beta$ suggests that the disorder becomes relevant when the dilution
$d>0.1$, which leads to the absence of self-averaging in thermodynamic
quantities such as the magnetic susceptibility and specific heat at
temperatures in the region around the critical point \cite{aharony1996}.
$\chi$ and $C$ are related to the fluctuations in the system's magnetization
and energy, respectively. Without self-averaging, a statistical fit to the
$\chi$ and $C$ data cannot be done with high confidence because values would
fall within a range that has a standard deviation that does not diminish
with an increase in the system size or the number of disorder realizations.
Even with $100$ realizations of the disorder, we find that fitting the data
to the scaling and finite-size scaling relations results in critical
exponent values with very low correlation coefficients.

\section{Summary and conclusion}
\label{sec:conclusion}

The interplay between disorder, interactions, and thermal fluctuations can
be explored in site-diluted Ising model in two dimensions. In this work, we
determine the magnetization $M$ per site, total energy $E$ per site, magnetic
susceptibility $\chi$ per site, and specific heat $C$ per site while
varying the dilution concentration $d$ and temperature $T$ in site-diluted
Ising systems on the square lattice. We are able to locate a critical
temperature $T_c$ and critical dilution concentration $d_c$ where the
phase transition occurs. We find that the estimated critical temperature
$T_c$ decreases linearly as the amount of dilution $d$ is increased. Our
data on how $\beta$ depends on $d$ suggests a transition in critical
behavior from Strong Universality at low dilutions, $d<0.1$, to Weak
Universality at high dilutions, $d>0.1$. Lastly, we find a wide and
nonsingular increase in magnetic susceptibility at low temperatures and
high dilutions region.

\begin{acknowledgments}
  The author is grateful to J.R.K. Bautista, V.R.L. Accad, M.A.D. Navarro,
  and L.P.M. Vargas for insightful discussions.
\end{acknowledgments}

\bibliography{myreferences}

\end{document}